\documentclass[twocolumn,amsmath,amssymb,nobibnotes,aps,prl,superscriptaddress,reprint]{revtex4-1}

\usepackage{microtype}
\usepackage{mathtools}
\usepackage{array}
\usepackage{bbm}
\usepackage{amsthm}

\newtheorem*{thm}{Theorem}

\usepackage{hyperref}

\usepackage{amsmath,amssymb,amsfonts,bbm,graphicx,hyperref,color,times}
\usepackage{subfigure}
\usepackage[english]{babel}

\newcommand{\bra}[1]{\ensuremath{\langle #1|}}
\newcommand{\ket}[1]{\ensuremath{|#1 \rangle}}

\newcommand{\ketbra}[2]{\ensuremath{| #1 \rangle\hspace{-2pt} \langle #2 |}}

\newcommand{\eref}[1]{(\ref{#1})}
\newcommand{\fref}[1]{figure \ref{#1}}
\newcommand{\Fref}[1]{Figure \ref{#1}}

\newcommand{\llrr}[1]{\ensuremath{\left( #1\right)}}
\newcommand{\llrrq}[1]{\ensuremath{\left[ #1\right]}}

\DeclareMathOperator{\Tr}{\hbox{Tr}}

\begin{document}

\author{D. Tamascelli}
\affiliation{Universit{\`a} degli Studi di Milano, Dipartimento di Fisica, Via Celoria 16, I-20133 Milano, Italy}
\affiliation{Institute of Theoretical Physics, Universit{\"a}t Ulm, Albert-Einstein-Allee 11D-89069 Ulm, Germany}
\author{A. Smirne}
\affiliation{Institute of Theoretical Physics, Universit{\"a}t Ulm, Albert-Einstein-Allee 11D-89069 Ulm, Germany}
\author{S.~F. Huelga}
\affiliation{Institute of Theoretical Physics, Universit{\"a}t Ulm, Albert-Einstein-Allee 11D-89069 Ulm, Germany}
\author{M.~B. Plenio}
\affiliation{Institute of Theoretical Physics, Universit{\"a}t Ulm, Albert-Einstein-Allee 11D-89069 Ulm, Germany}
\title{Non-perturbative treatment of non-Markovian dynamics of open quantum systems}
\begin{abstract}
   We identify the conditions that guarantee equivalence of the reduced dynamics of an open quantum system (OQS) for two different types of environments -- one a continuous bosonic environment leading to a unitary system-environment evolution and the other a discrete-mode bosonic environment resulting in a
system-mode (non-unitary) Lindbladian evolution.  Assuming initial Gaussian states for the environments, we prove that the two OQS dynamics are equivalent if both the expectation values and two-time correlation functions of the environmental interaction operators are the same at all times for the two configurations. Since the numerical and analytical description of a discrete-mode environment undergoing a Lindbladian evolution is significantly more efficient than that of a continuous bosonic environment {in a} unitary evolution, our result represents a powerful, non-perturbative tool to describe complex and  possibly highly non-Markovian dynamics. As a special application, we recover and generalize the well-known pseudomodes approach to open system dynamics.
\end{abstract}

\maketitle

\textit{Introduction.---}
Each and every realisation of physical quantum systems
will unavoidably suffer from interactions with uncontrollable degrees of freedom, namely the
surrounding environment. In certain situations it is possible to model such interaction as
resulting in white noise, amenable to an effective description in terms of a Lindblad master
equation \cite{Carmichael1993,Breuer2002,Gardiner2004,Rivas2012}. 
In general, however, the relevant noise sources
    originating from the interaction with structured environments give rise to non-Markovian
    effects and call for a more complex characterization \cite{Rivas2014,Breuer2016,deVega2017}. 
    Solid-state implementations of qubits \cite{Metha2016,Wu2016,Gambetta2017},
    nanoscale quantum thermal machines \cite{Esposito2015,Uzdin2016},  
    sensing and metrology \cite{Jones2009,Chin2011},
    energy-charge conversion and exciton transport in
solid-state devices \cite{Ribeiro2015,Mitchison2017}, or biological light har\-vesting complexes
\cite{Huelga2000,Pelzer2013}, are typical instances in which deviations from a Lindbladian
evolution can play a significant role.

The simulation of even simple OQSs interacting
with structured environments is a formidable task. If one adopts state-of-the-art numerical
methods \cite{Tanimura1989,Makarov1994,Prior2010,Tama2015} 
for the simulation of the dynamics of a quantum system,
 only systems of a few qubits are accessible. 
 A similar situation arises in proposed quantum physical simulators of system-environment
 interaction \cite{Porras2008}.
The difficulty, in both cases, is due to the large
number of environmental degrees of freedom 
affecting the reduced dynamics of the OQS under investigation. 
Several approaches have been developed
to map the original model into a unitarily equivalent one, which is easier to deal with, e.g.,
because it possesses a more suitable configuration for the application of proper numerical techniques 
\cite{Garg1985,Prior2010,Martinazzo2011,Strasberg2016,Newman2017}.
Yet, the simulation of such equivalent unitary models remains challenging, 
since the number of environmental degrees of freedom
involved is essentially unchanged.

\begin{figure}[t]
 \includegraphics[width=0.95\columnwidth]{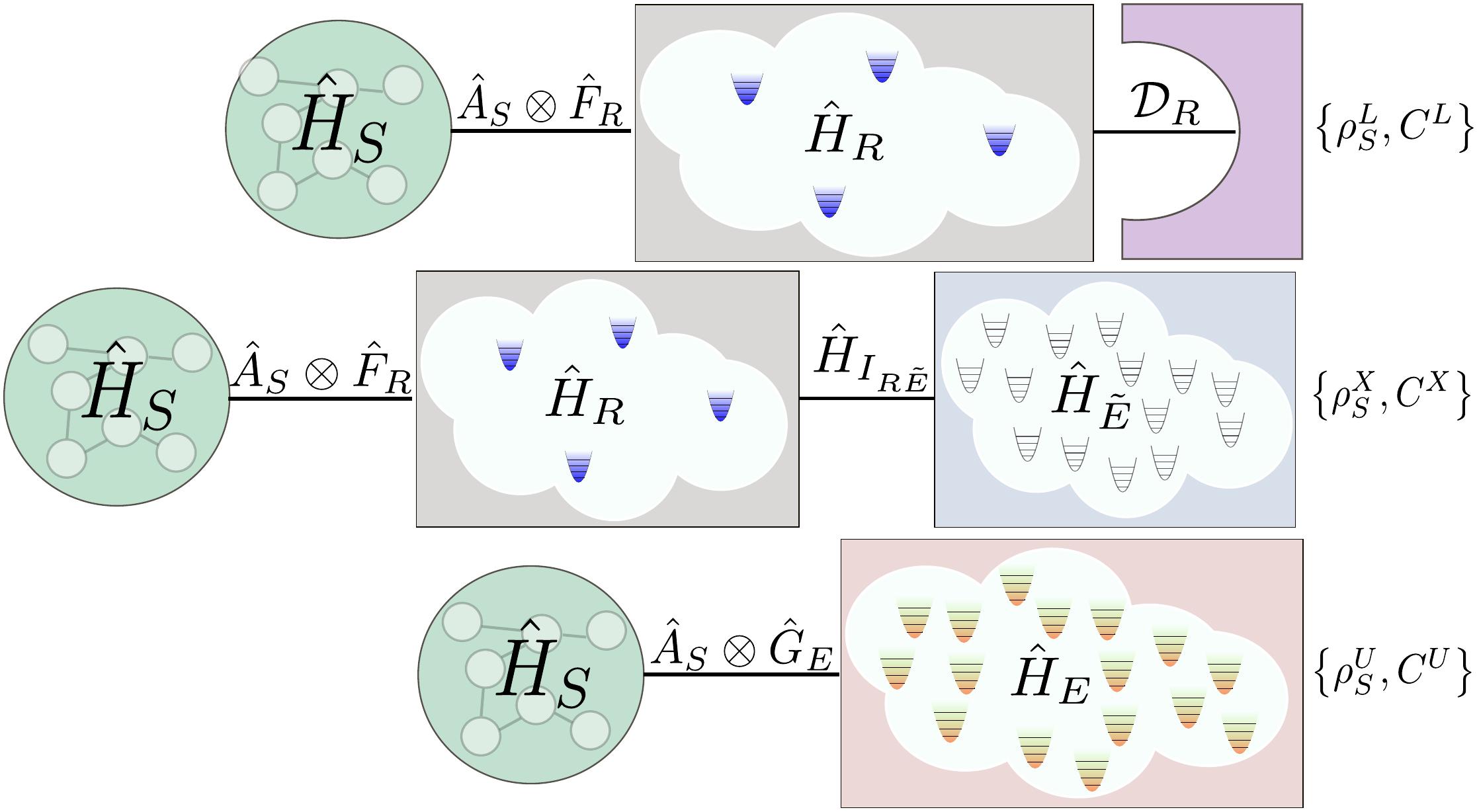}
\caption{Graphical representation of the considered configurations. Upper diagram: A possibly
composite system
    interacting with the environment $R$ undergoing the Lindblad dynamics as defined in Eq. \eref{eq:lin}.
    Lower: Same system interacting with the unitarily evolving environment $E$ (see Eq. 
    \eref{eq:hse}). Middle: The environment $R$ has been extended as to include additional modes with free
    evolution $H_{\tilde{E}}$  and interacting with the $R$ modes through $\hat{V}_{R\tilde{E}}$ (see Eq. \eref{eq:hx}).\label{fig:models}}
\end{figure}

On the other hand, if one is actually interested in the evolution of the open system only,
it is clear that the problem would be simplified drastically by
finding simpler auxiliary systems, which might not be
directly related to the original ones at the level of the overall dynamics, 
but which yield the same reduced dynamics for the open system.
In this regard, a powerful idea is to decompose the action of the environment into a non-Markovian core which, in turn, interacts with a Markovian environment.
The former interacts coherently with the open system and encloses all the memory effects during the evolution, 
while the latter can be characterized effectively by a Lindblad equation and represents the residual
unidirectional leaking of information out of the non-Markovian core
\cite{Imamoglu1994,Roden2011,Dzhioev2011,Ajisaka2012,Mostame2012,Schwarz2016,Lemmer2017} 
(see Fig.\ref{fig:models}). 
Of course, the most appealing feature of this approach is that the resulting configuration will be generally
much simpler than the original unitary one, having to deal with
a considerably smaller number of degrees of freedom.  

The possibility to reproduce the reduced OQS dynamics obtained from a unitary 
evolution involving a complex environment via a simpler environment which itself is subject to a Lindblad dynamics
is usually 
supported by a good agreement with experimental data or numerical analysis, 
as well as by approximative arguments, which can be applied 
in certain specific regimes. Nevertheless, rigorous results or theorems of some generality are still lacking and the only exact result
was derived in \cite{Garraway1997}, for the spin-boson model,
with a specific form of the interaction and a zero-temperature environment. There, a procedure was introduced
to replace the environment with infinitely many degrees of freedom by
a finite set of auxiliary harmonic modes, the so-called pseudomodes, 
which proved to be a very useful tool to characterize open-quantum-system dynamics \cite{Ban1998,Dalton2001,Maniscalco2008,Li2009,Mazzola2009,Fanchini2010,Rodin2011,Man2012,Zhang2012,Schonleber2015}.

In this work, we provide a proof of the general equivalence between
the reduced dynamics of an OQS 
interacting unitarily with a bosonic environment
and the dynamics of the same OQS interacting with a typically much simpler
harmonic environment
subject to a Lindblad evolution.
We prove that, for initial Gausssian states of the environments, the equivalence
is guaranteed if the environmental expectation values and two-time correlation functions of the two
configurations are equal for all times. 
We stress that, while this is well-known if one compares the reduced dynamics of two unitary
evolutions \cite{Feynmann1963,vanKampen1974,Breuer2002,Gasbarri2017},
it is \emph{a-priori} not obvious that 
the same still applies when comparing two reduced dynamics obtained from a unitary
and a non-unitary, Lindbladian evolution. 
The result holds irrespective of the strength of the system-environment interaction
or the structure of the environment,
thus providing a general non-perturbative way to describe
OQS dynamics, possibly highly non-Markovian ones. 
As a special case, we recover the
equivalence between the reduced dynamics of the spin-boson model and the description
given by the pseuodomodes, directly generalizing it to
different forms of the coupling.

\textit{The main result.---}
We start  by introducing the non-unitary configuration, which consists of a quantum system $S$ interacting with a bosonic environment $R$,
which is, in turn, subject to a Lindblad evolution. The system-environment Hamiltonian
reads
\begin{align}
\hat{H}_{SR} = \hat{H}_S +  \hat{H}_R + \sum^{\kappa}_{j=1}
\hat{A}_{S, j} \otimes \hat{F}_{R,j},
\label{eq:hamL}
\end{align}
{where, for a system of dimension $d_S$, $\kappa$ can take values in $1,2,\ldots,d_S^2$.}
Here and in what follows we imply the tensor product with the identity, so that $\hat{H}_S$ will be used instead of
$\hat{H}_S \otimes \mathbbm{1}$, and so on. The Lindbladian dynamics of the bipartite system
$S-R$ is fixed by the master equation ($\hbar = 1$)
\begin{equation}
    \dot{\rho}_{SR}(t) = \mathcal{L}_{SR}\llrrq{\rho_{SR}(t)} = -i \llrrq{\hat{H}_{SR},\rho_{SR}(t)} + \mathcal{D}_R\llrrq{\rho_{SR}(t)},
    \label{eq:lin}
\end{equation}
where
\begin{align}
    \mathcal{D}_{R} \llrrq{\rho} = \sum_{j=1}^\ell \gamma_j \llrr{ \hat{L}_{R,j} \rho \hat{L}_{R,j}^\dagger
    -\frac{1}{2} \left \{ \hat{L}_{R,j}^\dagger \hat{L}_{R,j}, \rho \right \}}
    \label{eq:dissLind}
\end{align}
acts {on $R$ only and $\ell$ determines the number of degrees of freedom of
$R$ and hence the complexity of the non-unitary model.}
We consider {time-independent arbitrary Lindblad operators
$\hat{L}_{R,j}$ and coefficients} $\gamma_j \geq 0$, ensuring the complete
positivity of the evolution \cite{Lindblad1976,Breuer2002}, and a factorized initial state
$\rho_{SR}(0) = \rho_S(0) \otimes \rho_R(0).$
We denote by $\rho_S^L(t)$ the corresponding reduced state of
the system $S$ at time $t$, namely:
\begin{align}
    \rho_S^L(t) = \Tr_R\left \{ e^ {\mathcal{L}_{SR} t} \llrrq{\rho_S(0) \otimes \rho_R(0)}\right \}.
    \label{eq:redLind}
\end{align} 
Here, $F_{R,j}(t)$ are the expectation values of the $R$ interaction operators
with respect to the ``free'' evolution
of the environment $R$ (i.e., without taking into account the presence of the system $S$),
$F_{R,j}(t)=\Tr_R \left\{\hat{F}_{R,j} e^{\mathcal{L}_R t}\left[\rho_{R}(0)\right]\right\}$,
where $\mathcal{L}_R$ is the generator
\begin{equation}\label{eq:ldag}
\mathcal{L}_R [\rho] = - i \left[\hat{H}_R, \rho \right] + \mathcal{D}_R[\rho ].
\end{equation}
Moreover, we denote by
\begin{align}
    C_{j j'}^L(t+s,s) = \Tr_R \left\{\hat{F}_{R,j}e^{\mathcal{L}_{R}t}\left[ \hat{F}_{R,j'} e^{\mathcal{L}_R s}\left[
\rho_{R}(0)\right]\right]\right\}
\label{eq:cljj}
\end{align}
the environment two-time correlation functions.

The other configuration we consider is given by the same system $S$ interacting unitarily with a
{bosonic}
environment $E$. The unitary $S-E$ evolution is determined by the Hamiltonian 
\begin{equation}\label{eq:hse}
\hat{H}_{SE} = \hat{H}_S+\hat{H}_E  + \sum^{\kappa}_{j=1} \hat{A}_{S, j} \otimes \hat{G}_{E,j},
\end{equation}
and the initial  factorized state $\rho_{SE}(0) = \rho_S(0) \otimes \rho_E(0)$.
We denote with $\rho_S^U(t)$ the corresponding $S$ reduced state at time $t$, i.e.,
\begin{equation}
\rho_S^U(t) = \Tr_E\left\{e^{-i \hat{H}_{SE} t}\left(\rho_S(0) \otimes \rho_E(0)\right) e^{i
    \hat{H}_{SE} t}\right\},
\end{equation}
the expectation values of the interaction terms as
$G_{E,j}(t)=\Tr_E \left\{\hat{G}_{E,j} e^{-i \hat{H}_{E} t}\rho_{E}(0)e^{i \hat{H}_{E} t}\right\}$ 
and with $C^U_{j j'}(t+s,s)$ the two-time correlation functions
\begin{eqnarray}
C^U_{j j'}(t+s,s) & = \Tr_E\left \{e^{i \hat{H}_{E} (t+s)}\hat{G}_{E,j}e^{-i \hat{H}_{E} (t+s)} \right
. \nonumber \\
&\left . e^{i \hat{H}_{E} s} \hat{G}_{E,j'} e^{-i \hat{H}_{E} s} \rho_E(0)\right\}. \label{eq:corru}
\end{eqnarray}

\begin{thm} \label{th:1}
    Given the two systems described by Eqs. \eqref{eq:hamL}-\eqref{eq:cljj} and Eqs.
    \eqref{eq:hse}-\eqref{eq:corru} respectively, if both $\rho_R(0)$ and $\rho_E(0)$ are Gaussian states, then the following implication holds:
\begin{align}
\left.
  \begin{tabular}{ccc}
$F_{R,j}(t)$& $=$ &$G_{E,j}(t)$ \\
$C^L_{j j'}(t+s,s)$  & $=$ & $C^U_{j j'}(t+s,s)$
  \end{tabular}
\right\}&
\,\,\, \forall j, j', t,s\geq 0 \nonumber \\
&\hspace{-3cm} \Longrightarrow \quad\rho_S^L(t) = \rho_S^U(t) \,\,\, \forall t. 
\end{align}
\end{thm}
{\it Sketch of the Proof.} The proof relies on two main steps, which we
represent as separate lemmas. 
Lemma 1 establishes a dilation \cite{Hudson1984,Parthasarathy1992,Gardiner2004,Barchielli2013} 
of the Lindblad equation on the bipartite system $S-R$ in Eq.\eqref{eq:lin}
to a unitary evolution on a tripartite system $S-R-\tilde{E}$,
such that the \emph{exact} reduced dynamics of $S$ under the latter is equal to that under the Lindbladian dynamics on $S-R$.
 Because of that, we can translate the comparison between
the two open-system dynamics of the configurations we are interested in into a
comparison between
the open-system dynamics of $S$ obtained from the unitaries on
$S-E$ and $S-R-\tilde{E}$, respectively.
These two open-system dynamics can be easily shown to be equivalent if \emph{their} environmental 
expectation values and
two-time correlation functions are equal (as we show in the last part of the proof).
But now Lemma 2 shows that the two-time correlation functions
of operators on $R$ with respect to the unitary dynamics on $R-\tilde{E}$
are equal to those obtained via the reduced Lindbladian dynamical maps on $R$
\cite{sideRem},
i.e., to $C^L_{j j'}(t+s,s)$, while the analogous
correspondence between the expectation values is implied directly by Lemma 1. Hence, we get 
the implication stated in the Theorem.
 \Fref{fig:models} represents the different configurations involved in the proofs of the Lemmas
 and the Theorem,
which are given in \cite{supplement}.
\\[5pt]\textbf{Lemma 1.} Consider a tripartite system $S-R-\tilde{E}$ undergoing the unitary evolution fixed by
\begin{align}
\hat{H}_{SR\tilde{E}} &= \hat{H}_{SR}+\hat{H}_{\tilde{E}}  + \hat{V}_{R\tilde{E}}, \nonumber\\
\hat{H}_{\tilde{E}} &=  \sum^{\ell}_{j=1} \int_{-\infty}^{\infty} d \omega\,  \omega \,  
\hat{b}^{\dag}_{\tilde{E}}(\omega, j) \hat{b}_{\tilde{E}}(\omega, j), \label{eq:hx}\\
\hat{V}_{R\tilde{E}} &= \sum^{\ell}_{j=1} \sqrt{-\frac{\gamma_j}{2 \pi}}
\int_{-\infty}^{\infty}\hspace{-5pt} d \omega 
\hat{L}_{R,j} \hat{b}^{\dag}_{\tilde{E}}(\omega, j) - \hat{L}^{\dag}_{R,j}
\hat{b}_{\tilde{E}}(\omega, j), \nonumber 
\end{align}
where $\hat{b}_{\tilde{E}}(\omega, j)$ and $\hat{b}^{\dag}_{\tilde{E}}(\omega, j)$ are bosonic
annihilation and creation operators of a Fock space $\mathcal{H}_{\tilde{E},j}$,
$\left[\hat{b}_{\tilde{E}}(\omega, j), \hat{b}^{\dag}_{\tilde{E}}(\omega', j')\right] = \delta_{j
j'} \delta(\omega-\omega')$,
%
and the global Fock space associated with the environment $\tilde{E}$
is indeed the tensor product $\mathcal{H}_{\tilde{E}} = \otimes_{j=1}^\ell \mathcal{H}_{\tilde{E},j}$.
Let the initial state be
$
\rho_{SR\tilde{E}}(0) = \rho_{SR}(0) \otimes \ketbra{0_{\tilde{E}}}{0_{\tilde{E}}},
$
where $\ket{0_{\tilde{E}}} =  \otimes_{j=1}^\ell  \ket{0_j}$ is the vacuum state of $\mathcal{H}_{\tilde{E}}$. 
Moreover, denote as $\rho_{SR}^X(t)$ the reduced $S-R$ state at time $t$, i.e.,
\begin{equation}\label{eq:rhox}
\rho_{SR}^X(t) = \Tr_{\tilde{E}}\left\{e^{-i \hat{H}_{SR\tilde{E}}  t}\left(\rho_{SR}(0) \otimes
\ketbra{0_{\tilde{E}}}{0_{\tilde{E}}}\right) e^{i \hat{H}_{SR\tilde{E}}  t}\right\};
\end{equation}
then (still denoting with $\rho_{SR}(t)$ the state fulfilling Eq.\eqref{eq:lin})
\begin{equation}\label{eq:step1}
\rho_{SR}^X(t) = \rho_{SR}(t).
\end{equation}
\textbf{Lemma 2.} 
Given the unitary dynamics on $R-\tilde{E}$ fixed by the Hamiltonian 
$\hat{H}_{R\tilde{E}} = \hat{H}_{R}+\hat{H}_{\tilde{E}} +\hat{V}_{R\tilde{E}}$, 
see Eq.\eqref{eq:hx}, its correlation functions
\begin{align}
C^X_{j j'}(t+s,s) &= \Tr_{RE}\left\{e^{i \hat{H}_{R\tilde{E}} (t+s)}\hat{F}_{R,j}e^{-i
    \hat{H}_{R\tilde{E}}(t+s)} \right . \\
    & \left . e^{i \hat{H}_{R\tilde{E}}s}  \hat{F}_{R,j'}e^{-i \hat{H}_{R\tilde{E}}s} (\rho_R(0)\otimes\rho_{\tilde{E}}(0))\right\}\nonumber 
\end{align}
satisfy
\begin{equation}\label{eq:step2}
C^X_{j j'}(t+s,s) =C^L_{j j'}(t+s,s) \quad \forall t,s \geq 0.
\end{equation}

\vspace{-3pt}
The Theorem shows that if we fix the open system $S$,
i.e., its free dynamics as given by $\hat{H}_S$
and how it interacts with the environment as given by the $\hat{A}_{S, j}$,
the equivalence of the expectation values $F_{R,j}(t)$ and $G_{E,j}(t)$ and of the correlation functions 
$C^L_{j j'}(t+s,s)$ and $C^U_{j j'}(t+s,s)$ ensures the equivalence of the open-system dynamics
of the two configurations.
On the other hand, the environmental Hamiltonians $\hat{H}_{R}$
and $\hat{H}_{E}$ need not to be equal,
neither do
the environmental interaction operators $\hat{F}_{R,j}$ and $\hat{G}_{E,j}$ 
or the initial Gaussian states $\rho_{R}(0)$ and $\rho_{E}(0)$,
nor we are setting \emph{a-priori} any constraint on the form of the Lindblad operators $\hat{L}_{R,j}$.

Crucially, the equivalence between the two reduced dynamics,  governed by Eqs. \eqref{eq:hamL}-\eqref{eq:cljj} and Eqs.
    \eqref{eq:hse}-\eqref{eq:corru} respectively, can be then guaranteed 
in the presence of a Lindbladian environment $R$ which is much simpler than the unitary one $E$,
being characterized by a considerably smaller number of degrees of freedom.
We will give an explicit example in the next paragraph. We also emphasize that
the Theorem provides us with a manifestly non-perturbative way to deal
with general, non-Markovian open-system dynamics, 
since we have not set 
any restriction on the strength of the coupling, nor on the 
structure of the environment.
In addition, our result can be straightforwardly generalized beyond the assumption
of initial Gaussian states, by asking for the equality of the higher order 
correlation functions of the environments $R$ and $E$, see \cite{supplement}.
Finally, we stress that our results in their present form establish the
equivalence 
between the reduced dynamics of the unitary and the Lindbladian scenarios
for 
single-time expectation values of open-system observables while the case of 
multi-time correlators remains open.


\textit{The spin-boson model and the pseuodomodes.---}
%
Consider a two-level system linearly interacting with an environment of harmonic oscillators via a
Hamiltonian as in Eq.\eqref{eq:hse} \cite{Leggett1987,Breuer2002}.
We use $\sigma_i$, $i=x,y,z$ to denote the Pauli matrices, 
 $\sigma_+ = \sigma^{\dag}_-=\ketbra{1}{0}$, with $\ket{1},\ket{0}$ eigenvectors of $\sigma_z$, 
while $\hat{a}_{\omega}$
and $\hat{a}^{\dag}_{\omega}$ are the annihilation and creation operators of the bosonic field,
$[\hat{a}_{\omega}, \hat{a}^\dagger_{\omega'}] = \delta(\omega-\omega')$.
In particular, we take  
$\hat{H}_S=\omega\sigma_z$ and $\hat{H}_E = \int_{-\infty}^{+\infty} d\omega \, \omega \,  \hat{a}_\omega^\dagger
    \hat{a}_\omega$, while we consider two different forms of interaction.
In one case we set
$\hat{A}_S = \sigma_x$ and 
$\hat{G}_E = \int^{\infty}_{-\infty} d \omega (g(\omega) \hat{a}_\omega+g^*(\omega)\hat{a}^{\dagger}_\omega)$,
while in the other we take two terms in the interaction
(we use primed letters to denote the corresponding interaction operators)
$\hat{A'}_{S,1} =\sigma_-$, $\hat{A'}_{S,2} =\sigma_+$, 
$\hat{G'}_{E,1} = \int^{\infty}_{-\infty} d \omega g(\omega) \hat{a}_\omega$
and $\hat{G'}_{E,2}=\hat{G'}^{\dagger}_{E,1}$.
Note that the second form of the coupling can be obtained from the former after the rotating wave approximation 
and, importantly, it conserves the total number of excitations \cite{Breuer2002,Bina2012}.
Restricting for simplicity to a zero-temperature environment, 
$\rho_E(0) = \ketbra{0}{0}$, which is stationary with respect to $\hat{H}_E$, 
the expectation values of the environmental interaction operators vanish and
there is only one non-trivial
two-time correlation function for both versions of the system-environment coupling;
namely,
\begin{align}
    {C_{SB}^{U}(t)} & = \Tr_E \left\{e^{i \hat{H}_E t}\hat{G'}_{E,1} e^{-i \hat{H}_E t}\hat{G'}_{E,2} \ketbra{0}{0}\right\} \nonumber\\
& =\int^{\infty}_{-\infty} d \omega \mathcal{S}(\omega) e^{-i \omega t}, \label{eq:csbu}
\end{align}
where we introduced $\mathcal{S}(\omega) = |g(\omega)|^2$, usually referred to as correlation
spectrum {\cite{sideRemSD}}.

Let us now consider the non-unitary dynamics fixed by Eqs. \eqref{eq:hamL}-\eqref{eq:dissLind},
with the same $\hat{H}_S$ and $\hat{A}_{S,j}$ specified above,
and where $R$ is defined via a set of independent auxiliary harmonic modes,
with
the annihilation and creation operators $\hat{c}_j$
and $\hat{c}^\dagger_j$, $j=1,\ldots \ell$, such that $[\hat{c}_j, \hat{c}^\dagger_l ] = \delta_{jl}$.
Furthermore, we
set $\hat{H}_R=\sum_{j=1}^\ell \eta_j \hat{c}^\dagger_j \hat{c}_j$ and we still consider two forms of the coupling, 
$\hat{F}_{R} = \sum^\ell_{j=1} (\lambda\hat{c}_j+\lambda^*\hat{c}^{\dagger}_j)$ 
(together with $\hat{A}_S = \sigma_x$) or 
$\hat{F'}_{R,1} =\sum^\ell_{j=1} \lambda\hat{c}_j$ and $\hat{F'}_{R,2} =\hat{F'}^\dagger_{R,1}$
(together with $\hat{A'}_{S,1}=\hat{A'}^\dagger_{S,2} = \sigma_+$) .
The Lindblad generator acting on $R$ is defined as in Eq.\eqref{eq:ldag},
with dissipator
\begin{align} 
\mathcal{D}_R[\rho] & = \sum_{j=1}^\ell \gamma_j 
\llrr{\hat{c}_j \rho\hat{c}_j^\dagger -\frac{1}{2} \left \{\hat{c}_j^\dagger \hat{c}_j,
    \rho \right \}}. \label{eq:exampleD}  
\end{align}
Fixing $\rho_R(0)=\ketbra{0}{0}$, 
which is stationary with respect to $\mathcal{L}_R$,
for both the forms
of the coupling the environmental
expectation values vanish and there is still only one non-trivial
two-time correlation function, see Eq.\eqref{eq:cljj},
given by
\begin{align}
    {C_{SB}^L(t)} & = \Tr_R \left\{\hat{F'}_{R,1}e^{\mathcal{L}_{R}t}\left[ \hat{F'}_{R,2} \ketbra{0}{0}\right]\right\} \nonumber\\
& =|\lambda|^2 \sum^\ell_{j=1} e^{(i \eta_j -\gamma_j/2)t}. \label{eq:csbl}
\end{align}
We conclude that for any unitary spin-boson dynamics such that 
the Fourier transform of the spectrum $S(\omega)$, see Eq.\eqref{eq:csbu},
can be written as a sum of $\ell$ exponentials as in Eq. \eqref{eq:csbl},
we can define an equivalent non-unitary dynamics, which only involves $\ell$
modes and which yields the same reduced dynamics on the two-level system $S$ at any time.
For the excitation-preserving form of the coupling,
this is equivalent to the result obtained 
in \cite{Garraway1997} and the $\ell$ modes of the environment $R$
precisely identify with the \emph{pseudomodes} introduced there.
Hence, not only  have we recovered this result as a direct application of our 
Theorem, but we have also generalized it to a different
form of the system-environment coupling, not preserving the number of excitations.

\begin{figure}
    \includegraphics[width=0.8\columnwidth]{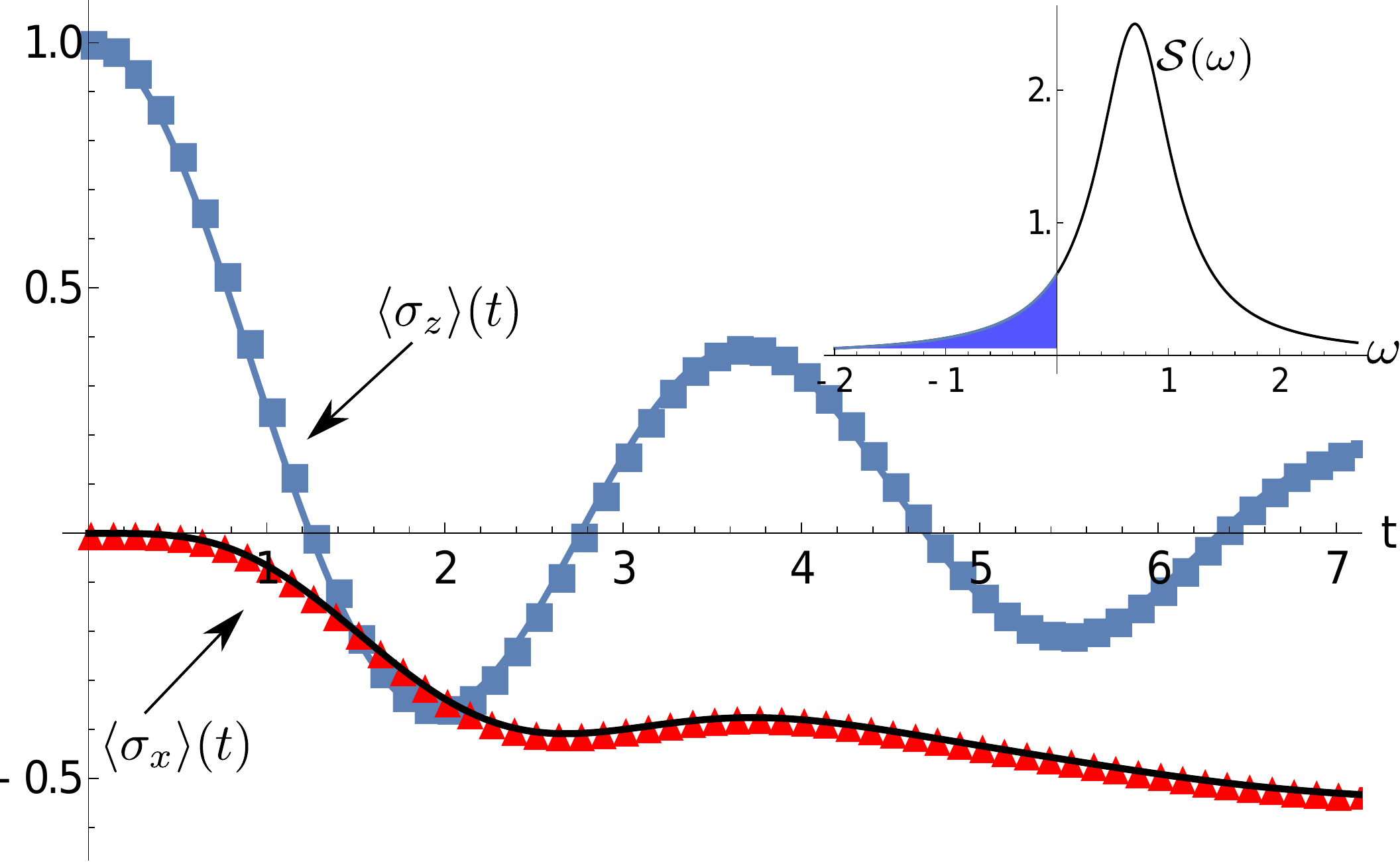}
   \caption{ \label{fig:tedopa} Comparison between the expectation values of the observables
   $\sigma_z$ and $\sigma_x$ on the reduced state $\rho_S(t)$
as obtained by Lindblad (solid lines) and unitary (TEDOPA) dynamics (blue squares and red
triangles).
The parameters are $\omega= \eta_1 = 0.7$, $\lambda=0.6$,
$\gamma_1=0.4$. The Lorentzian correlation spectrum $\mathcal{S}(\omega)$ for this parametrization is shown in the
inset.}
\end{figure}

To give an explicit example, let us consider the Lindblad equation for a single harmonic oscillator,
interacting via the excitation non-preserving coupling with the two-level system and damped as in Eq.\eqref{eq:exampleD},
with $\ell = 1$. 
For any choice of the defining parameters $\lambda, \eta_1$ and $\gamma_1$ the reduced dynamics
of the qubit interacting with the damped harmonic oscillator will be the same as the reduced
dynamics of a qubit interacting with infinitely many harmonic oscillators,
with correlation spectrum given by the Lorentzian 
$\mathcal{S}(\omega)= |\lambda|^2 \gamma_1/(\gamma_1^2 + (\omega - \eta_1)^2)$, see Eq.\eqref{eq:csbl}.
In \fref{fig:tedopa} we show an example of these two reduced dynamics of the two-level system. 
In particular, the parameters have been chosen as to have the mode at
resonance with the qubit ($\eta_1 = \omega$) and the
strong coupling regime  ($\lambda > \gamma_1$). {Note that such regime can be 
    realized in an ion-trap
setup \cite{Lemmer2017}.}
Moreover, the
part of the correlation spectrum on the negative frequencies (which is fixed by the ratio $\gamma_1/\eta_1$)
is not negligible, which
provides us with a regime where the approximative argument put forward
in \cite{Imamoglu1994} does not apply. 
The unitary evolution has been simulated using the numerically exact
TEDOPA algorithm \cite{Prior2010}. The two curves in the graphs overlap perfectly. 
This exemplifies the equivalence between an environment with a Lorentzian
spectrum and an environment consisting of a damped harmonic oscillator in a setting
where neither the original pseudomode approach \cite{Garraway1997}, 
nor the approximated argument of \cite{Imamoglu1994} would apply.


The simple example now described allows us to emphasize another important issue. 
If
we want to restrict the definition of the environment to 
the positive frequencies only, we set (for both forms of the coupling) $g(\omega)= 0$
for $\omega<0$,
and hence $\mathcal{S}(\omega) = 0$
for $\omega<0$. The latter condition can be also easily seen to follow from
the fluctuation-dissipation relation \cite{Breuer2002} applied to a zero-temperature environment. 
Now, no finite number of auxiliary modes can reproduce exactly a two-time
correlation function, whose  inverse Fourier transform vanishes on an interval of the real axis, see Eq.\eqref{eq:csbl}.
Hence, features commonly exploited to characterize
unitary systems, {such as the fluctuation-dissipation relation itself (see also
    \cite{Talkner1986}),}
might impose
qualitative differences between unitary and Lindbladian correlation functions, thus questioning the general usefulness of our result.
However, also in these situations one can take advantage of the constructive nature of the proof
of the Theorem by formulating a Lindbladian dynamics such that
$C^U(t) \approx C^L(t)$. The construction of Lemmas 1 and Lemma 2 can still  
be pursued, so that one would end up with two unitary dynamics with similar correlation functions,
$C^X(t) \approx C^U(t)$.
By applying suitable bounding procedures which apply to unitary dynamics \cite{Mascherpa2017},
we could then provide a precise bound to the difference between the reduced dynamics
$\rho^U(t)$ and $\rho^L(t)$ based on the difference between their correlation functions.

\textit{Conclusions.---}
In this work, we proved
in rigorous terms conditions that ensure the equivalence
between the two reduced dynamics of an open system as resulting from, respectively, an overall unitary dynamics
and the interaction with a simpler environment undergoing a Lindblad evolution.
This yields a general non-perturbative way to characterize
even highly non-Markovian dynamics with a smaller
set of degrees of freedom.

Our result paves the way for rigorous investigations
of the validity of the simulation of a unitary dynamics with a computationally simpler Lindblad
evolution, possibly dealing with an approximated equality of the two.
From a more abstract perspective, our result also 
represents an extension of the input-output formalism \cite{Gardiner2004}
to the bipartite scenario, where the system subjected to the white noise consists
of the open system and its non-Markovian core.
Indeed, it will be of interest to investigate how this relates to other
non-Markovian input-output approaches introduced in the literature \cite{Diosi2012,Zhang2013,Xue2017}.
{In addition, the fact that an input/output formalism can also be developed for a
    fermionic bath \cite{Gardiner2004b} suggests that our result can be
    extended to this domain. We plan to study such an extension, as well as to treat multi-time
    correlation functions of the open-system observables, in order to fully characterize the
system's spectral response.}
%
%
%
%

\acknowledgments{We thank Andreas Lemmer for many useful discussions; we acknowledge support by the state of
Baden-W\"urttemberg through bwHPC and the German Research Foundation (DFG) through grant no INST
40/467-1 FUGG; financial support by the ERC Synergy grant BioQ and the EU project QUCHIP is acknowledged.}

\setcounter{equation}{0}
\renewcommand{\theequation}{S\arabic{equation}}     
\onecolumngrid
\section{Supplemental material}
\section{Lemma 1}
\begin{proof}
The Heisenberg equation of motion for a generic $S-R$ operator $\hat{O}_{SR}$ induced by the
Hamiltonian in Eq.(11) of the main text  reads [see Eq. (5.3.36) of \cite{Gardiner2004}]
\begin{align}
\frac{d}{d t} \hat{O}_{SR}(t) = & i \left[\hat{H}_{SR}(t), \hat{O}_{SR}(t)\right] -\sum_{j=1}^{\ell}\left[\hat{O}_{SR}(t),\hat{L}^{\dag}_{R,j}(t)\right]\left(\frac{\gamma_j}{2} \hat{L}_{R,j}(t)+\sqrt{\gamma_j} \hat{b}_{in}(t, j) \right) \nonumber\\
&+\sum_{j=1}^{\ell}\left(\frac{\gamma_j}{2} \hat{L}^{\dag}_{R,j}(t)+\sqrt{\gamma_j}
\hat{b}^{\dag}_{in}(t, j) \right) \left[\hat{O}_{SR}(t),\hat{L}_{R,j}(t)\right], \label{eq:539} 
\end{align}
where, indeed, the time-dependence in $\hat{H}_{SR}(t)$ and $\hat{L}_{R,j}(t)$ is due to the Heisenberg picture,
\begin{equation}\label{eq:heis}
\hat{H}_{SR}(t) = e^{i \hat{H}_{SR\tilde{E}} t} \hat{H}_{SR} e^{-i \hat{H}_{SR\tilde{E}}t}
\end{equation}
(and the same for $\hat{L}_{R,j}(t)$);
it is worth emphasizing that, while the operator at the initial time $\hat{O}_{SR}$
is non-trivial only on $\mathcal{H}_{S} \otimes \mathcal{H}_R$, this is not necessarily the case for the operator at time $t$, 
$\hat{O}_{SR}(t)$, which will act non-trivially also on $\mathcal{H}_{\tilde{E}}$.
Finally, we have introduced the operator $\hat{b}_{in}(t, j)$ on $\mathcal{H}_{\tilde{E}, j}$ as
\begin{equation}
\hat{b}_{in}(t, j) = \frac{1}{\sqrt{2 \pi}} \int_{-\infty}^{\infty} d \omega e^{- i \omega t} \hat{b}_{\tilde{E}}(\omega, j)
\end{equation}
(and the same for $\hat{b}^{\dag}_{in}(t, j)$);
we stress that the dependence on time is included parametrically in the definition of $\hat{b}_{in}(t, j)$,
which, for any time $t$, is an operator on $\mathcal{H}_{\tilde{E}}$ only.
Note that this is nothing else than the so-called ``input field'', which characterizes the input-output formalism \cite{Gardiner2004}
and satisfies the commutation relations
\begin{align}\label{eq:inou}
&\left[\hat{b}_{in}(t,j), \hat{b}^{\dag}_{in}(s,j')\right] = \delta_{j j'}\delta(t-s), \\ 
&\left[\hat{b}_{in}(t,j), \hat{b}_{in}(s,j')\right] = 0 \quad \forall t,s \geq 0.
\end{align}

Taking the expectation value, with the notation 
\begin{align}
\left\langle  \hat{O}_{SR} (t) \right\rangle_{SR\tilde{E}} = \Tr_{SR\tilde{E}}\left\{\hat{O}_{SR}(t) \left(\rho_{SR}(0) \otimes \ketbra{0_{\tilde{E}}}{0_{\tilde{E}}}\right) \right\}, \label{eq:supsup}
\end{align}
we get
\begin{align}
\frac{d}{d t} \langle \hat{O}_{SR}(t)\rangle_{SR\tilde{E}} &=\left\langle i
    \left[\hat{H}_{SR}(t), \hat{O}_{SR}(t)\right] 
 +\sum_{j=1}^{\ell}\gamma_j\left(\hat{L}^{\dag}_{R,j}(t)\hat{O}_{SR}(t) \hat{L}_{R,j}(t)
- \frac{1}{2}\left\{\hat{L}_{R,j}(t) \hat{L}^{\dag}_{R,j}(t),  \hat{O}_{SR}(t)\right\}\right) \right
. \nonumber\\
& \left . +\sum_{j=1}^{\ell} \sqrt{\gamma_j} \left(\left[\hat{O}_{SR}(t),\hat{L}^{\dag}_{R,j}(t)\right]
\hat{b}_{in}(t, j) 
 + \hat{b}^{\dag}_{in}(t, j) \left[\hat{O}_{SR}(t),\hat{L}_{R,j}(t)\right]\right)
\right\rangle_{SR\tilde{E}} . \label{eq:supsup2}
\end{align}
But then, the contribution from the terms in the second line is equal to zero since 
\begin{equation}\label{eq:zer}
\hat{b}_{in}(t, j) \ket{0} = 0 = \bra{0}\hat{b}^{\dag}_{in}(t, j).
\end{equation}
Hence, going back to the Schr{\"o}dinger picture, see Eq.\eqref{eq:heis},
and using the definition of reduced dynamics in Eq.(12) of the main text
both in the left hand side and in the right hand side of Eq.\eqref{eq:supsup2}
(see also Eq.\eqref{eq:supsup}),
we end up with
\begin{align}
\frac{d}{d t} Tr_{SR}\left\{ \hat{O}_{SR} \rho^X_{SR}(t)\right\}= \Tr_{SR}\left\{ \hat{O}_{SR} \left (- i
    \left[\hat{H}_{SR}, \rho^X_{SR}(t)\right]  +\sum_{j=1}^{\ell}\gamma_j\left(\hat{L}_{R,j}\rho^X_{SR}(t) \hat{L}^{\dag}_{R,j}
-\frac{1}{2}\left\{\hat{L}^{\dag}_{R,j}(t) \hat{L}_{R,j}(t),  \rho^X_{SR}(t)\right\}\right) \right )
\right\}.
\end{align}
Since this holds for any $S-R$ operator, we can conclude that the statistical operator
$\rho^X_{SR}(t)$ satisfies the Lindblad equation (2) in the main text
and then Eq.(13) in the main text holds.\\
\end{proof}

Note that the Hamiltonian $\hat{V}_{R\tilde{E}}$ defined in Eq.(11) of the main text
is singular, due to the flat distribution of the couplings. However, as usual within the input-output formalism, a proper
unitary evolution on a Hilbert space can be formulated by means of the quantum stochastic
calculus, for which the reader is referred to \cite{Hudson1984,Parthasarathy1992,Barchielli2013}.

Finally, the previous Lemma holds for any initial condition on the $S-R$ system; in
    particular, in the Theorem
we use it for $\rho_{SR}(0) = \rho_S(0) \otimes \rho_R(0)$ since we 
assume an initial $S-R$ product state.
\section{Lemma 2}
\begin{proof}
Using the Heisenberg picture with respect to $\hat{H}_{R\tilde{E}}$ on two generic operators of $R$, $\hat{M}_R$ and $\hat{N}_R$, one has
indeed [compare with Eq.\eqref{eq:539}]
    \begin{align}
\frac{d}{d t} \hat{M}_R(t+s) \hat{N}_R(s) &= \left(i \left[\hat{H}_R(t+s), \hat{M}_R(t+s)\right] 
-\sum_{j=1}^{\ell}\left[\hat{M}_{R}(t+s),\hat{L}^{\dag}_{R,j}(t+s)\right]\left(\frac{\gamma_j}{2} \hat{L}_{R,j}(t+s)+\sqrt{\gamma_j} \hat{b}_{in}(t+s, j) \right) \right. \nonumber\\
&+\left.\sum_{j=1}^{\ell}\left(\frac{\gamma_j}{2} \hat{L}^{\dag}_{R,j}(t+s)+\sqrt{\gamma_j} \hat{b}^{\dag}_{in}(t+s, j) \right) \left[\hat{M}_{R}(t+s),\hat{L}_{R,j}(t+s)\right]
\right) \hat{N}_R(s).
\end{align}
Now, the crucial point is that \cite{Gardiner2004}
\begin{equation}\label{eq:comm}
\left[\hat{b}_{in}(t',j), \hat{N}_R(t)\right] = 0 \quad \forall t'>t,
\end{equation}
since the solution of the Heisenberg equation of motion for $\hat{N}_R(t)$  (see
Eq.\eqref{eq:539}) will depend on $\hat{N}_R(s)$, as well as on $\hat{b}_{in}(s,l)$ at previous times, $s< t$, but will not depend
on the future values of the input field, $\hat{b}_{in}(t',j)$ with $t'>t$. This, together with
Eq.\eqref{eq:inou}, implies Eq.\eqref{eq:comm}.
Hence, evaluating the derivative of the correlation function by taking the trace on $R-\tilde{E}$ of the previous expression applied
to the initial state $\rho_R(0) \otimes \ketbra{0}{0}$, using Eqs. \eqref{eq:zer} and \eqref{eq:comm} and the notation
\begin{align}
\left\langle \hat{M}_{R}(t) \right\rangle_{R\tilde{E}} = \Tr_{R\tilde{E}} \left\{\hat{M}_R(t)(\rho_R(0) \otimes \ketbra{0}{0}) \right\},
\end{align}
we end up with
\begin{align}
&\frac{d}{d t} \left\langle \hat{M}_R(t+s) \hat{N}_R(s) \right\rangle_{R\tilde{E}} 
 =
\left\langle  \left[i \left[\hat{H}_R(t+s), \hat{M}_R(t+s)\right] \right.\right. \nonumber \\
&\left.\left.+\sum_{j=1}^{\ell} \gamma_j \left(\hat{L}^{\dag}_{R,j}(t+s) \hat{M}_{R}(t+s) \hat{L}_{R,j}(t+s) - \frac{1}{2}\left\{\hat{L}_{R,j}(t+s)\hat{L}^{\dag}_{R,j}(t+s), \hat{M}_{R}(t+s)\right\}
\right)\right] \hat{N}_R(s)\right\rangle_{R\tilde{E}}  \nonumber \\
&= \Tr_{R\tilde{E}} \left\{ \mathcal{L}^{\dag}_R[ \hat{M}_R] e^{-i \hat{H}_{R \tilde{E}}(t+s)} \hat{N}_R(s)  (\rho_R(0) \otimes \ketbra{0}{0}) e^{i \hat{H}_{R \tilde{E}}(t+s)}\right\},  \label{eq:eqq}
\end{align}
where we used the adjoint of the generator in Eq.(5) in the main text, which is defined via the duality relation
\begin{align}
\Tr_R\left\{\hat{O}_R \mathcal{L}_R [\rho_R]\right\} = \Tr_R\left\{\mathcal{L}^{\dag}_R[\hat{O}_R] \rho_R\right\} 
\end{align}
for any state and bounded operator on $\mathcal{H}_R$; the derivative of the correlation function has been defined by continuity at $t=0$.
Moreover, the initial condition is, by definition,
\begin{equation}\label{eq:icby}
\left\langle \hat{M}_R(s) \hat{N}_R(s) \right\rangle_{R\tilde{E}} = \Tr_{R\tilde{E}} \left\{e^{i \hat{H}_{R \tilde{E}}s} \hat{M}_R \hat{N}_R e^{-i \hat{H}_{R \tilde{E}}s}(\rho_R(0) \otimes \ketbra{0}{0})\right\} = 
\Tr_{R} \left\{\hat{M}_R \hat{N}_R e^{\mathcal{L}_R s}[\rho_R(0)]\right\},
\end{equation}
where, in the second equality, we used that the reduced dynamics on $R+\tilde{E}$ is exactly given 
by the generator in Eq.(5) of the main text [which is a special case of Lemma 1, when the system S is trivial].
Now, consider the basis of operators on $R$, $\left\{\hat{M}_{a b, R} = \ketbra{\psi_a}{\psi_b}\right\}_{a,b}$, 
where $\left\{\ket{\psi_a}\right\}_{a}$ is a basis of the Hilbert space $\mathcal{H}_R$,
so that one has \cite{Carmichael1993}
\begin{equation}\label{eq:lmat}
\mathcal{L}^{\dag}_R\left[\hat{M}_{a b, R} \right] = \sum_{c d} W_{ab}^{cd} \hat{M}_{c d, R} \quad  W_{ab}^{cd} = \bra{\psi_c}\mathcal{L}^{\dag}_R\left[\ketbra{\psi_{a}}{\psi_b}\right] \ket{\psi_d}.
\end{equation}
Looking at the corresponding two-time correlations functions, 
Eqs.\eqref{eq:eqq} and \eqref{eq:lmat} imply
\begin{equation}
\frac{d}{d t} \left\langle \hat{M}_{a b,R}(t+s) \hat{N}_R(s) \right\rangle
= \sum_{c d} W_{ab}^{cd} \left\langle\hat{M}_{c d,R}(t+s) \hat{N}_R(s) \right\rangle.
\end{equation}
Using once more Eq.\eqref{eq:lmat}, it is easy to see that
$$
\left\langle \hat{M}_{a b,R}(t+s) \hat{N}_R(s) \right\rangle = \Tr_R\left\{e^{\mathcal{L}^{\dag}_R t}[ \hat{M}_{a b,R}] \hat{N}_R e^{\mathcal{L}_R s}\rho_R(0)\right\}
$$
solves the previous system of equations, with initial condition given by Eq.\eqref{eq:icby}.
Since this is the case for all the elements on a basis of operators in $\mathcal{H}_R$, we can write an analogous relation by replacing $\hat{M}_{a b,R}$ with $\hat{F}_{R,j}$;
identifying $\hat{N}_R$ with $\hat{F}_{R,j'}$ and using the definition of adjoint generator, we thus
get Eq.(15) of the main text.\\
\end{proof}

\section{Theorem}
\begin{proof}
{The Theorem directly follows from the fact that 
given a unitary dynamics, fixed by an Hamiltonian 
$\hat{H}_{S\bar{E}} = \hat{H}_S+\hat{H}_{\bar{E}}  + \sum^{\kappa}_{j=1} \hat{A}_{S, j} \otimes
\hat{K}_{\bar{E},j}$ 
and with an initial product global state, $\rho_{S\bar{E}}(0)=\rho_{S}(0)\otimes\rho_{\bar{E}}(0)$, 
the reduced dynamics of $S$ is fixed \emph{in a unique way} by i) the free Hamiltonian of the system $\hat{H}_{S}$; ii) the system interaction terms $\hat{A}_{S, j}$ and iii) 
the time-ordered multi-time environment correlation functions
\begin{equation}\label{eq:multi}
C_{\bar{E}}(t_1,\ldots t_k)= \Tr_{\bar{E}}\left\{\hat{K}_{\bar{E}, j_1}(t_1)\ldots \hat{K}_{\bar{E}, j_k}(t_k) \rho_{\bar{E}}(0)\right\},
\end{equation}
where $\hat{K}_{\bar{E}, j}(t) = e^{i \hat{H}_{\bar{E}} t}\hat{K}_{\bar{E}, j}e^{-i
    \hat{H}_{\bar{E}} t}$ and $j_1,\ldots,j_k$ assume values between $1$ and $\kappa$. This can be seen
    from the partial trace of the Dyson expansion, or other perturbative expansions, such as
    the time-convolutionless projection operator method \cite{Feynmann1963,vanKampen1974,Breuer2002,Gasbarri2017}. 
Now, if the initial state of the environment $\rho_{\bar{E}}(0)$ is Gaussian, the multi-time correlation functions $C_{\bar{E}}(t_1,\ldots t_k)$ can be fully expressed in terms of 
the expectation values $K_{\bar{E}, j}(t) = \Tr_{\bar{E}}\left\{\hat{K}_{\bar{E}, j}(t) \rho_{\bar{E}}(0)\right\}$ and
the two-time correlation functions
$C_{j j'}(t,t') = \Tr_{\bar{E}}\left\{\hat{K}_{\bar{E}, j}(t) \hat{K}_{\bar{E}, j'}(t')
\rho_{\bar{E}}(0)\right\}$, with $t>t'$.}

Hence, consider the unitary dynamics fixed by the Hamiltonian $\hat{H}_{SE}$ in Eq.(7) of the main
text.
and the unitary dynamics fixed by  $\hat{H}_{SR\tilde{E}}$ in Eq.(11) of the main text, along with the corresponding reduced dynamics on $S$;
moreover, consider
the expectation value of the interaction operators $\hat{F}_{R,j}$
$$
F^X_{R,j}(t)=
\Tr_{\tilde{E}}\left\{\hat{F}_{R,j} e^{-i \hat{H}_{R\tilde{E}}  t}\left(\rho_{R}(0) \otimes \ketbra{0}{0}\right)e^{i \hat{H}_{R\tilde{E}}  t} \right\}.
$$
Since in both cases the free Hamiltonian $\hat{H}_S$, as well as the 
interaction operators $\hat{A}_{S,j}$ are the same, the environments are bosonic
and the initial conditions are given by
product states with Gaussian states of the environments, we conclude that,
if $F^X_{R,j}(t) = G_{E,j}(t)$ and $C^X_{jj'}(t,t')=C^U_{jj'}(t,t')$ for any $t\geq t'$, then $\rho^X(t)=\rho^U(t)$ for any $t$.

The Theorem directly follows from the two previous lemmas, since $\rho^X_S(t)=\rho^L_S(t)$
and $F^X_{R,j}(t) =F_{R,j}(t)$
because of Lemma~1 (the second equality for Lemma~1 in the case of a trivial system $S$),
while $C^X_{jj'}(t,t')=C^L_{jj'}(t,t')$ for $t\geq t'$ because of Lemma~2.
 
\end{proof}

Indeed, we can generalize our result beyond the assumption of initial Gaussian states
$\rho_R(0)$ and $\rho_E(0)$ by asking that also the higher order correlations functions are equal:
the equivalence between the two reduced dynamics would still follow from Lemma 1,
Lemma 2 ({generalized
to the case of environmental multi-time correlation functions, i.e., replacing
$\hat{N}_R(s)$ with different operators evaluated at different, decreasing times}) and what we have just proved, simply without expressing the multi-time
correlation functions, see Eq.\eqref{eq:multi}, in terms of the expectation values and two-time correlation functions.
{While the control over all the higher order environmental correlation function
might be unrealistic in practical situations, 
our Theorem provides some useful hints to
deal with the simplified description of a complex open-quantum-system dynamics
via an auxiliary system, also beyond the assumption
of initial Gaussian states, if one is satisfied with an approximated equivalence between the two reduced
dynamics. The argument goes along the same lines as the analogous one made in the main text after the example
represented in Figure 2.
We know that, e.g., in the TCL perturbative expansion \cite{Breuer2002} the $n-th$
order correlation functions appear in the equations of motion only starting from the $n-th$ order term
of the expansion. Hence, relying on the Lemmas 1 and 2, we can guarantee
that if the first $m$ correlation functions are equal, the difference 
between the two reduced dynamics will be smaller or equal to the error made 
by neglecting the terms from the order $m+1$ up (which can be estimated
with techniques applied to the perturbative expansions of unitary global dynamics \cite{Breuer2002,Mascherpa2017}).  }

As a final remark let us stress that, in general, the assumption of having equal 
expectation values for the interaction operators,
$F_{R,j}(t) = G_{E,j}(t)$,
cannot be relaxed, even if the initial state of the environments are stationary
(i.e., $\mathcal{L}_R[\rho_R(0)]=0$ and $[\hat{H}_E, \rho_E(0)]=0$). 
It is in fact true that in this case $F_{R,j}(t)=F_{R,j}(0)$
and $G_{E,j}(t)=G_{E,j}(0)$ are time-independent and can be set equal to 0 by redefining the
Hamiltonians $\hat{H}_{SE}$ and $\hat{H}_{SR}$, but this would
imply a redefinition of the free Hamiltonian $\hat{H}_S$ itself, which might be different
in the two cases.
Explicitly, let us consider first the unitary $S-E$ dynamics:
one can define 
new interaction operators and a new free Hamiltonian  \cite{Rivas2012},
$$
\hat{G^\circ}_{E,j} = \hat{G}_{E,j} - \Tr_E\left\{\hat{G}_{E,j} \rho_E(0)\right\} \hat{\mathbbm{1}}
\qquad \hat{H^\circ}_S =  \hat{H}_S+\sum_{j=1}^\ell  \Tr_E\left\{\hat{G}_{E,j} \rho_E(0)\right\} \hat{A}_{S,j},
$$
without changing the global Hamiltonian, and then the reduced dynamics, but such that now the expectation
values of the environment interaction operators are zero at any time $t$.
Exactly the same procedure can be followed for the Hamiltonian $\hat{H}_{SR}$, which, along with the dissipator
$\mathcal{D}_R$, fixes the non-unitary dynamics we considered (see Eqs.(1)-(4) in the main text). 
However, it is clear that this would lead to the definition of a free Hamiltonian, 
$$
\hat{H^\diamond}_S =  \hat{H}_S+\sum_{j=1}^\ell  \Tr_R\left\{\hat{F}_{R,j} \rho_R(0)\right\} \hat{A}_{S,j},
$$
which is in general different from $\hat{H^\circ}_S$ if 
$\Tr_R\left\{\hat{F}_{R,j} \rho_R(0)\right\}\neq  \Tr_E\left\{\hat{G}_{E,j} \rho_E(0)\right\}$,
i.e., if $F_{R,j}(0) \neq G_{E,j}(0)$.
Of course, if $F_{R,j}(0) = G_{E,j}(0)$ one can eliminate the dependence on the expectation values in
the equations of motion of both the configurations 
via an identical redefinition of the interaction operators and free system Hamiltonian.

\end{document}